\documentclass[twocolumn,runningheads,natbib]{svjour2}

\smartqed  
\usepackage{graphicx}
\usepackage{deluxetable}
\journalname{Astrophysics and Space Science}
\begin{document}

\def\farcs{\hbox{$.\!\!^{\prime\prime}$}}
\def\farcm{\hbox{$.\!\!^{\prime}$}}
\def\cxo{{\em Chandra}}
\def\snr{supernova remnant}
\def\lesssim{\mathrel{\hbox{\rlap{\hbox{\lower4pt\hbox{$\sim$}}}\hbox{$<$}}}}
\def\gtrsim{\mathrel{\hbox{\rlap{\hbox{\lower4pt\hbox{$\sim$}}}\hbox{$>$}}}}
\def\ref{\par \hangindent 5 pt\noindent}
\def\gs{\lower 2pt \hbox{$\;\scriptscriptstyle \buildrel>\over\sim\;$}}

\title{UV emission from young and middle-aged pulsars
} \subtitle{Connecting X-rays with the optical}

\author{Oleg Kargaltsev         \and
        George Pavlov
}

\institute{O. Kargaltsev \at
              Pennsylvania State University
              525 Davey Lab, University Park, PA 16802 \\
              \email{oyk100@psu.edu}           
           \and
            G. Pavlov \at
           Pennsylvania State University
              525 Davey Lab, University Park, PA 16802 \\
              \email{pavlov@astro.psu.edu}
}

\date{Received: date / Accepted: date}

\maketitle

\begin{abstract}
We present the UV spectroscopy and timing of three nearby
 pulsars (Vela, B0656+14 and Geminga)
 recently observed with the Space Telescope Imaging Spectrograph.
 We also review the
optical and X-ray properties of these pulsars and establish their
connection with the UV properties. We show that the multiwavelengths
properties of neutron stars (NSs) vary significantly within the
sample of middle-aged pulsars. Even
larger differences are found between
 the thermal components of Ge-minga and B0656+14 as compared to those of
 radio-quiet isolated NSs. These differences could be attributed to
different properties of the NS surface layers.

\keywords{Neutron Stars \and Pulsars \and Geminga \and PSR~B0656+14 \and Vela pulsar
 \and RX~J1856.5--3754 \and RX~J0720.4--3125 \and RX~J1308.6+2127}
\PACS{97.60.Gb \and 97.60.Jd }
\end{abstract}

\section{Introduction}
\label{intro}

Optical through X-rays radiation from a typical isolated neutron
star (NS) is expected to exhibit two
 components: thermal radiation from the NS surface and
non-thermal radiation from the NS magnetosphere.
The spectrum of {\em non-thermal radiation}
can be described
 by a power-law (PL) model with
a spectral index $-1\lesssim\alpha\lesssim 0$ ($F_{\nu}\propto
\nu^{\alpha}$). This radiation is commonly interpreted as
synchrotron emission from relativistic electrons/positrons
accelerated by the electric fields near the NS surface and from
secondary particles produced in the pair cascades. The
magnetospheric radiation  is intrinsically anisotropic and,
therefore,
it shows strong pulsations.
The non-thermal emission
 completely dominates the
multiwavelength spectrum in very
young pulsars (e.g., Crab and B0540--69).

As the magnetospheric emission becomes fainter with increasing
pulsar age,
 {\em thermal emission}
becomes detect-able
at $\tau\equiv P/2\dot{P}\gtrsim 10$ kyrs, being seen as a ``thermal hump''
on top of the flat non-thermal spectrum.
This
 thermal radiation
 is
  emitted from the NS surface which can
 be non-uniformly heated (e.g., due to anisotropic heat conductivity
 of the crust and
bombardment by relativistic particles).
  Indeed, in several cases
 at least
  two thermal components with different temperatures
are need-ed to fit the X-ray spectrum (e.g., Pavlov et al.\ 2002;
Zavlin \& Pavlov 2004a; De Luca et al.\ 2005).
 The cooler {\em thermal soft} (TS) component is commonly interpreted as emission from
 the bulk of the NS surface.
The hotter
{\em thermal hard} (TH) component is often attributed to the NS
polar caps (PCs) which can be additionally heated by the
magnetospheric particles.
As expected, the TS temperatures generally decrease with
pulsar age. However, they do not fall onto a monotonically
decreasing cooling curve,
which suggests that NSs may have different masses
 (e.g., Yakovlev \& Pethick 2004).
In X-rays, the TS component usually shows weaker and broader
pulsations
than the TH component. The TS pulsations
can be
caused by
 non-uniformities of the NS surface temperature
and by anisotropy of
 local emissivity in the strong magnetic
field.

 Spectra of middle-aged pulsars ($10\lesssim\tau\lesssim500$ kyr)
are particulary interesting because they often exhibit all the three
emission components.
The nearby middle-aged pulsars, Vela ($\tau\approx11$ kyrs),
B0656+14 (hereafter B0656; $\tau\approx110$ kyrs) and Geminga
($\tau\approx340$ kyrs), are relatively bright and have been
extensively studied in both the optical and X-rays. Given their
different ages, the three pulsars provide a representative cut
through the population of middle-aged pulsars,
allowing one to
study the evolution of the pulsar properties.

\begin{table}[t]
\setlength{\tabcolsep}{0.85\tabcolsep}
\caption{Basic properties of three middle-aged pulsars and
three RQINSs
with measured spin-down parameters and/or parallaxes (Kaplan \& van
Kerkwijk 2005ab;
see also van Kerkwijk \& Kaplan, these proceedings).} \centering
\label{tab:1}       
\begin{tabular}{lcccccc}
\hline\noalign{\smallskip}
Name & Dist.\ & $~~\tau$ & $P$ & $\log\dot{E}$
& B & $r_{\rm pc}$ \\[3pt]
& $~~\rm{pc}$ &  kyr & s & ergs s$^{-1}$ & TG & m \\[3pt]
\tableheadseprule\noalign{\smallskip}

Geminga  & $\!\!\!\!\!\!\sim 200$   & ~340 & 0.24 & 34.52
& 1.6 & 440 \\
B0656+14 & $288^{+33}_{-27}$ & ~110 & 0.38 & 34.58
& 4.7 & 346\\
    Vela & $293^{+19}_{-17}$ & ~~11  & 0.09  & 36.84
& 3.8 & 717\\
RX~J0720 & $333^{+167}_{-83}$ & 1900 & ~~8.4 & 30.67
& 24 & ~~~7\\
RX~J1308 & $\!\!\!\!\!\!\sim700$ & 1500 & 10.3  & 30.60
& 34 & ~~~7\\
RX~J1856 & $161^{+17}_{-14}$ & $\!\!\sim$400 & ...  & ...  & ... & ...\\
\noalign{\smallskip}\hline
\end{tabular}\\
\end{table}

The X-ray spectra of these three pulsars can be described by a
three-component, TS+TH+PL, model, with the thermal components
modeled as BBs. However, such fits result in substantially different
model parameters in each case.

 The youngest Vela pulsar shows the highest
 $T_{\rm TS}\approx1.2$ MK emitted from a region with $R_{\rm TS}\approx3.7$ km.
This $R_{\rm TS}$ is substantially
smaller than a typical NS radius, which
might indicate that a large fraction of
the NS surface is too cold to be seen in X-rays\footnote{
Alternatively, it may mean that the BB description of the
thermal
spectrum is inadequate, and a more
realistic atmosphere model should be used.
For instance, the
Vela pulsar spectrum  fits equally well by a two-component,
NS atmosphere plus PL, model, which gives $T\approx0.7$ MK at $R=13$
km (Pavlov et al.\ 2001; see also
Kargaltsev 2004).}.
A factor of ten
older B0656 has a lower temperature, $T_{\rm TS}\approx0.7$ MK, and
a much larger emitting area, $R_{\rm TS}\approx12$ km,
close to that of a typical NS. The oldest of the three pulsars, Geminga, is
even colder, $T_{\rm TS}\approx0.5$ MK, and has the largest emitting
area ($R_{\rm TS}\approx13$ km).

 The PL component,
with $\alpha_{\rm X}\approx-1$, is the strongest in the Vela pulsar,
$L_{X,{\rm PL}}\approx3\times10^{31}$ erg s$^{-1}$
 (in 0.2$-$10 keV), becoming weaker in B0656 ($\alpha_{\rm X}\approx -0.5$;
  $L_{X,{\rm PL}}\approx1.4\times10^{31}$ erg s$^{-1}$), and much
  weaker in Geminga ($\alpha_{\rm X}\sim-0.6$;
  $L_{X,{\rm PL}}\approx2.2\times10^{30}$ erg s$^{-1}$). This PL
  component is typically interpreted as
  synchrotron emission produced by relativistic particles in the
  pulsar margentosphere.
%

  The strength
  of the TH component also appears to decrease
   with pulsar age.
   Being relatively strong in the Vela pulsar ($L_{{\rm TH}}\approx6.3\times10^{31}$ erg
   s$^{-1}$), it
   becomes
    weaker in B0656 ($L_{{\rm TH}}\approx3.5\times10^{31}$ erg
   s$^{-1}$) and decreases dramatically
    in Geminga ($L_{{\rm TH}}\approx4.4\times10^{29}$ erg
   s$^{-1}$). The sizes of the TH emission regions are close to the
   conventional PC radii [$r_{\rm pc} \equiv (2\pi R_{NS}^{3}/cP)^{1/2}$; see Table 1] in
   Vela and B0656, while in Geminga, $R_{\rm TH}\approx46$ m, is surprisingly small when compared to
   its $r_{\rm pc}=440$ m.
This may suggest that the surface of
   Geminga is more uniformly heated and the ``TH component''
   simply mimics
   a phase-dependent PL component\footnote{From a phase-resolved analysis of the XMM data,
Jackson \& Halpern (2005) conclude that the X-ray spectrum of
Geminga can be alternatively described by a BB+PL model with a
phase-dependent PL slope.}. However, regardless of the exact nature
of the TH and PL components, the soft X-ray (0.3$-$2 keV) spectra
of these three pulsars are clearly dominated by the thermal
emission.

Observing the TS radiation over a wide range of frequencies is
particulary important.
 X-ray observations
can only probe the Wein tail of the TS spectrum emitted from the NS
surface. This information is insufficient to reconstruct the overall
shape of the spectrum  which can deviate from the simple black-body
due to the presence of an atmosphere
 or nonuniformity of surface
temperature. Chemical composition of the atmosphere can dramatically
affect the X-ray spectrum (Romani 1987; Pavlov et al.\ 1995; Zavlin,
Pavlov \& Shibanov 1996; Zavlin \& Pavlov 2002), with light element
atmospheres leading to a large Wien excess. Any surface temperature
inhomogeneities will also complicate the spectrum, with hotter
regions being increasingly important at higher frequencies. For
these reasons, comparison of thermal X-ray emission with optical-UV
emission on the Rayleigh-Jeans (R-J) side of the thermal hump
 is
particularly valuable. The challenge here is that the non-thermal
magnetospheric emission becomes increasingly dominant at longer
wavelengths.

\begin{table*}[t]
\caption{STIS observations} \centering
\label{tab:1}       
\begin{tabular}{lllcccc}
\hline\noalign{\smallskip}
Pulsar & Date & Instrument & Exposure, s & $\Delta\lambda^{~\rm a}$, \AA & ${\mathcal F}^{~\rm b}$, cgs & $f_{p}^{~\rm c}$, \% \\[3pt]
\tableheadseprule\noalign{\smallskip}
Geminga  & 2002\,Feb\,27 & NUV-MAMA/F25SRF2 & 11367 & 1800-3000 & $1.7\times 10^{-15}$ & 40 \\
         & 2002\,Feb\,26 & FUV-MAMA/G140L & 10674   & 1155-1702 & $3.7\times 10^{-15}$ & 65 \\
B0656+14 & 2001\,Sep\,\,1   & NUV-MAMA/PRISM & \,\,\,6791    & 1790-2950 & $2.6\times 10^{-15}$ & 67\\
         & 2001\,Nov\,16 & NUV-MAMA/PRISM & 12761   & 1790-2950 & $2.6\times 10^{-15}$ & 89\\
         & 2004\,Jan\,20  & FUV-MAMA/G140L & \,\,\,4950    & 1153-1700 & $4.2\times 10^{-15}$ & 64\\
    Vela & 2002\,May\,28 & NUV-MAMA/F25SRF2 & \,\,\,2895  & 1800-3000 & $7.2\times 10^{-15}$ & 87 \\
         & 2002\,May\,28 & FUV-MAMA/G140L & \,\,\,3060    & 1153-1701 & $8.2\times 10^{-15}$ & 73 \\
\noalign{\smallskip}\hline
\end{tabular}\\
\tablecomments{$^{~\rm a}$ -- instrument+filter passband;
 $^{~\rm b}$ -- observed flux in the corresponding passband (for the NUV
 observations of B0656, the flux has been measured from the two
 observations combined);
$^{~\rm c}$ -- intrinsic (corrected for the background) pulsed
fraction.}
\end{table*}

\begin{figure}
\hspace{0.5cm}
  \includegraphics[width=3.2in,angle=0]{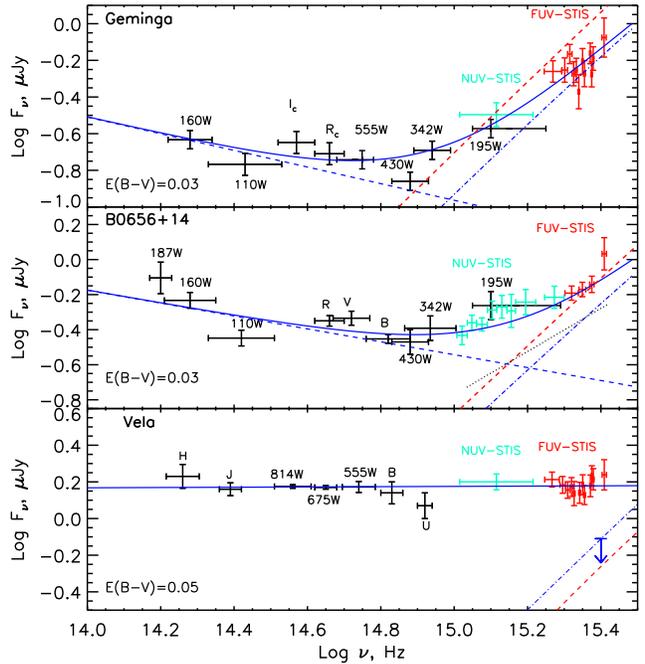}
\caption{Unabsorbed NIR through FUV spectra of the Geminga,
B0656 and Vela pulsars (top to bottom). The PL+BB fits are shown
by the solid blue lines. The contributions of the individual
components are also shown (blue dashed line for PL, blue dash-dotted
line for BB). The dashed red line shows the extrapolation of the TS
component fitted to the X-ray spectrum (see also Fig.~3). For
B0656 we have also plotted the fit to the NUV+FUV spectrum at the
pulse minima (black dotted line). }
\label{fig:1}       
\vspace{-0.5cm}
\end{figure}

The observations with ground-based telescopes and the Hubble Space
Telescope (HST) have provided multi-band photometry from near-IR
(NIR) to near-UV (NUV) and showed that
 the NIR-optical spectra of
the three
pulsars are non-thermal, with a hint of a
R-J component
seen in Geminga and B0656+14 at $\lambda\lesssim3000$ \AA.
 The observations in
 UBVRIHJ filters show that the optical spectrum of the Vela pulsar remains very
flat from NIR
 to UV, with $\alpha_{\rm O}\approx-0.1$
(Shibanov et al.\
 2003).
For the other two pulsars, the NIR-optical spectra are steeper,
$\alpha_{\rm O}\approx-0.4$ (Koptsevich et al.\ 2001; Shibanov et
al.\ 2006).

For all the three pulsars optical pulse profile measurements have
been carried out. The Vela and B0656, show strong non-sinusoidal
pulsations (Gouiff\'es 1997; Kern et al.\ 2003). Optical pulsations
of Geminga were only marginally detected in the B band (Shearer et
al.\ 1998).

Despite the extensive coverage in the optical and X-rays, very few
NUV observations had been carried out until recently, and the FUV
parts of the spectra have remained completely unexplored.
 To  fill this gap, we have undertaken an
observational campaign with the MAMA-NUV and MAMA-FUV detectors of
the Space Telescope Imaging Spectrograph (STIS) aboard the  HST.

\section{Observations with STIS in far- and near-UV}

We have observed all the there pulsars in the NUV (1800--3000 \AA)
and FUV (1150$-$1700 \AA) passbands (see Table~2). The data
reduction and analysis are described in detail by Kargaltsev et al.\
(2005), Romani et al.\ (2005), and Kargaltsev \& Pavlov (2006). Here
we summarize the main results of these observations.

\subsection{Geminga}

In the FUV passband we obtained a low-resolution spectrum of Geminga
using MAMA-FUV with the grating G140L. The observed flux in
the 1155--1702 \AA\ range is ${\mathcal F}_{\rm FUV}
= (3.72\pm 0.24) \times 10^{-15}$ ergs
s$^{-1}$ cm$^{-2}$, corresponding to the luminosity $L_{\rm
FUV}=4\pi d^2 {\mathcal F}_{\rm FUV} =(1.78\pm 0.11)\times 10^{28} d_{200}^2$
ergs s$^{-1}$. Fitting the spectrum with the absorbed PL model
 gives
$\alpha_{\rm FUV} = 1.43\pm 0.53$ for
 $E(B-V)=0.03$.  The PL slope is close to that of
the
R-J spectrum,
$F_\nu \propto \nu^2$,
suggesting that the observed radiation is dominated by thermal
emission from the NS surface. To estimate the NS surface
temperature, we fit the absorbed BB model to the
spectrum.
Since the FUV
frequencies are in the
R-J part of the spectrum,
the temperature values are strongly correlated with the
radius-to-distance ratio (approximately, $T\propto d^2/R^2$).
 For a typical
 NS radius, $R=13$ km, and the assumed distance $d=200$ pc,
the inferred
temperatures are
 $0.31\pm
0.01$
 and
$0.41\pm 0.02$ MK,
 for
$E(B-V)=0.03$ and 0.07, respectively.

The Geminga pulsar was imaged with MAMA-NUV using the broad-band
filter F25SRF2. The NUV flux
 depends on the
assumed spectral slope and extinction.
For a plausible E(B$-$V)=0.03
  and
$\alpha_{\rm NUV} =1$, the unabsorbed
  NUV flux is
  ${\mathcal F}_{\rm NUV} =2.2\times 10^{-15}$ ergs cm$^{-2}$ s$^{-1}$ in 1800$-$3000 \AA.

We used the 125 $\mu$s time resolution of STIS
for timing analysis of both NUV and FUV data.
We found statistically significant pulsations with frequencies
 lying within ($-$0.3,+0.7)\,$\mu$Hz around the  $f=4,217,608.6953\,
\mu{\rm Hz}$ frequency estimated from multi-epoch timing
observations in $\gamma$-rays and X-rays (Jackson et al.\ 2002).
The folded (source plus background) NUV light curve, plotted in
Figure 2, shows one broad (FWHM $\approx 0.8$ of the period),
flat-top peak per period, centered at the phase $\phi\approx 1.0$.
The most notable feature of the pulse profile is the narrow dip at
$\phi\approx 0.45$.
 The pulsed fraction, defined as the ratio of the number of
counts above the minimum level to the total number of counts in the
light curve, is about 28\%, which corresponds to the intrinsic
source pulsed fraction $f_{\rm p}\approx 40\%$.

 The
 FUV
  light curve, plotted in the same Figure 2,
also shows a sharp, asymmetric dip at
$\phi\approx0.45$ with an
 additional shallower dip
 at $\phi\approx 0.95$. The pulsed
fraction in the observed (source + background) radiation is about
 40\% (intrinsic pulsed fraction 60\%--70\%.).

\subsection{B0656+14}

The FUV spectrum of B0656 has been measured with MAMA-FUV using
the G140L grating. Unfortunately, only two
of the eight planned orbits of data
were collected because of the
 STIS failure on
August 3, 2004.
 The observed flux in the 1153--1700 \AA\ band
is
 ${\mathcal F}_{\rm FUV}= (4.3\pm 0.3) \times 10^{-15}$ ergs s$^{-1}$ cm$^{-2}$,
corresponding to
$L_{\rm FUV}=(4.2\pm 0.3)\times
10^{28} d_{288}^2$ ergs s$^{-1}$. The fit with the absorbed PL model
results in a rather steep spectral slope
$\alpha_{\rm FUV} = 1.51\pm 0.62$ for a
plausible $E(B-V)=0.03$. Similar to Geminga, this suggests that the
FUV
 radiation could be dominated by
thermal emission from the NS surface. The absorbed BB fit gives the
surface temperature of $0.71\pm 0.03$ MK for $E(B-V)=0.03$,  $R=13$
km, and the distance of $288$ pc (Brisken et al.\ 2003). The
corresponding unabsorbed bolometric luminosity is
 $L_{\rm bol}=(3.1\pm 0.5)\times 10^{32}$
ergs s$^{-1}$. Thus, the measured brightness temperature is
substantially higher than that of Geminga.

The NUV spectrum and light curve of B0656 have been obtained using
MAMA-NUV with PRISM (Shibanov et al.\ 2005). We re-analyzed these
data to facilitate the direct comparison with the FUV and other
multiwavelength data.
 The
observed flux in the 1790--2950 \AA\ range is
${\mathcal F}_{\rm NUV}= (2.63\pm 0.38) \times 10^{-15}$ erg s$^{-1}$ cm$^{-2}$,
corresponding to
$L_{\rm NUV}=(2.61\pm 0.38)\times
10^{28} d_{288}^2$ erg s$^{-1}$.
Fitting the spectrum with the
absorbed PL model,
we find
$\alpha_{\rm NUV} =1.09\pm 0.41$ for a plausible $E(B-V)=0.03$.
The PL slope is somewhat flatter than that
of the FUV spectrum, which can be explained by a larger
 contribution of the non-thermal component.

Making use of the event time-tags, we created FUV and NUV pulse
profiles folded with the radio ephemeris provided by M.\ Kramer
(2005; priv.\ comm.).
 The FUV
pulse profile (Fig.~2) shows two peaks of
approximately equal strengths.
The minimum at
$\phi\approx0.55$ is
  shallower than the
 other
minimum at $\phi\approx1.05$. The observed pulsed fraction,
$\approx36\%$, corresponds to the intrinsic puls-ed
fraction of $\approx64$\%. We also folded the light curves for each
of the NUV visits and co-added them into the combined NUV pulse
profile (Fig.\ 2).
The NUV pulse shape resembles the FUV one
but shows a higher
pulsed fraction (the
 observed and intrinsic pulsed fractions are 22\% and 80\%,
respectively). The lower pulsed fraction in FUV is evidence of a
larger contribution of thermal emission at the FUV wavelengths
 (thermal emission is intrinsically less pulsed than the non-thermal
emission). The NUV and FUV light curves can be interpreted as two
non-thermal peaks on top of a single low-amplitude thermal pulse
with the maximum at $\phi\approx0.5$. This interpretation would
imply a substantially softer spectrum at the pulse minima than at
the peaks.

To verify this assumption,
we produced the phase-resolved NUV through FUV spectrum for three
phase intervals $[0.5; 0.7]$, $[0.9; 1.2]$, and
$[0.2;0.5]+[0.7;0.9]$, corresponding to each of the two peaks and two
minima combined.
 The difference in the slopes of the
PL model fitted to the peaks spectra does not exceed the spectral
index uncertainties. However, the spectrum at the pulse minima is
substantially steeper
($\alpha =1.5\pm0.2$) than the peak
spectra
($\alpha= 0.6\pm0.2$).
This
behavior is in line with the expected larger
 contribution of thermal emission at the minima phases.

\subsection{Vela pulsar}

The spectroscopic (FUV) and imaging (NUV) data has been acquired
using exactly the same instrumental setup as in the Geminga observations
(see Table 1) and reduced in a similar way.

For a plausible PL slope, $\alpha_{\rm NUV} = 0$ (see \S3.3) and
color index $E(B-V)=0.05$ (estimated from the hydrogen column
density found from the X-ray fits; Sanwal et al.\ 2002), the
unabsorbed
  NUV flux
  is ${\mathcal F}_{\rm NUV} \approx 1.1\times 10^{-14}$ ergs cm$^{-2}$ s$^{-1}$ in 1800$-$3000 \AA.

 The FUV spectrum of the Vela pulsar is significantly flatter than those of Geminga
 and B0656, which reflects its predominantly non-thermal nature even throughout the FUV passband.
  The total
observed flux in the 1153--1701 \AA\ range is
 ${\mathcal F}_{\rm FUV}= (8.19\pm 0.36) \times 10^{-15}$ ergs s$^{-1}$ cm$^{-2}$,
corresponding to
$L_{\rm FUV}=(8.78\pm 0.39)\times 10^{28} $
ergs s$^{-1}$ at
$d=300$ pc (Dodson et al.\ 2003).
Fitting the spectrum with the absorbed power-law model,
we found
$\alpha_{\rm FUV} = 0.06\pm 0.39$
at $E(B-V) =0.05$.

Figure 2 shows the NUV and FUV light curves folded
with appropriate radio ephemeris (courtesy of R.\ N.\ Man-chester).   At
least four
 narrow peaks can be identified in each of the light curves.
 The
 higher signal-to-noise (S/N)
 NUV light curve
shows some substructure in the two main peaks. The observed (source
plus background) pulsed fractions
 are
$52\%$ and $73\%$ for the FUV and NUV light curves, respectively.
The corresponding intrinsic pulse fractions are 73\% and 87\%.

\begin{figure*}
\centering
  \includegraphics[width=0.7\textwidth]{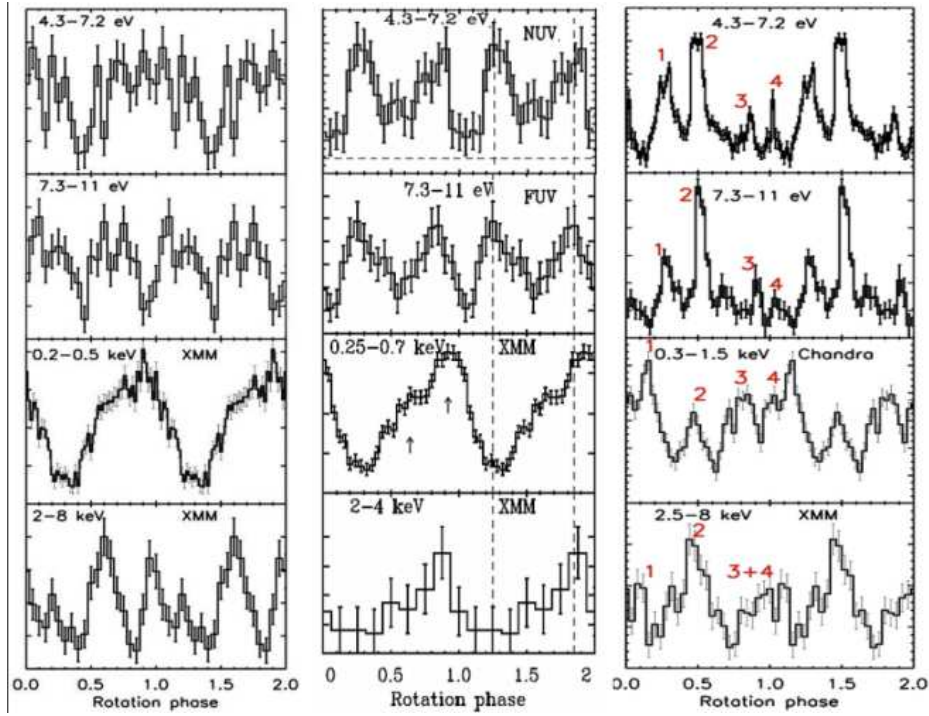}
\caption{NUV, FUV and X-ray (top to bottom) light curves of the
Gemigna, B0656+14 and Vela pulsars (left to right).}
\label{fig:2}       
\end{figure*}

\begin{figure*}
\centering
  \includegraphics[width=0.77\textwidth]{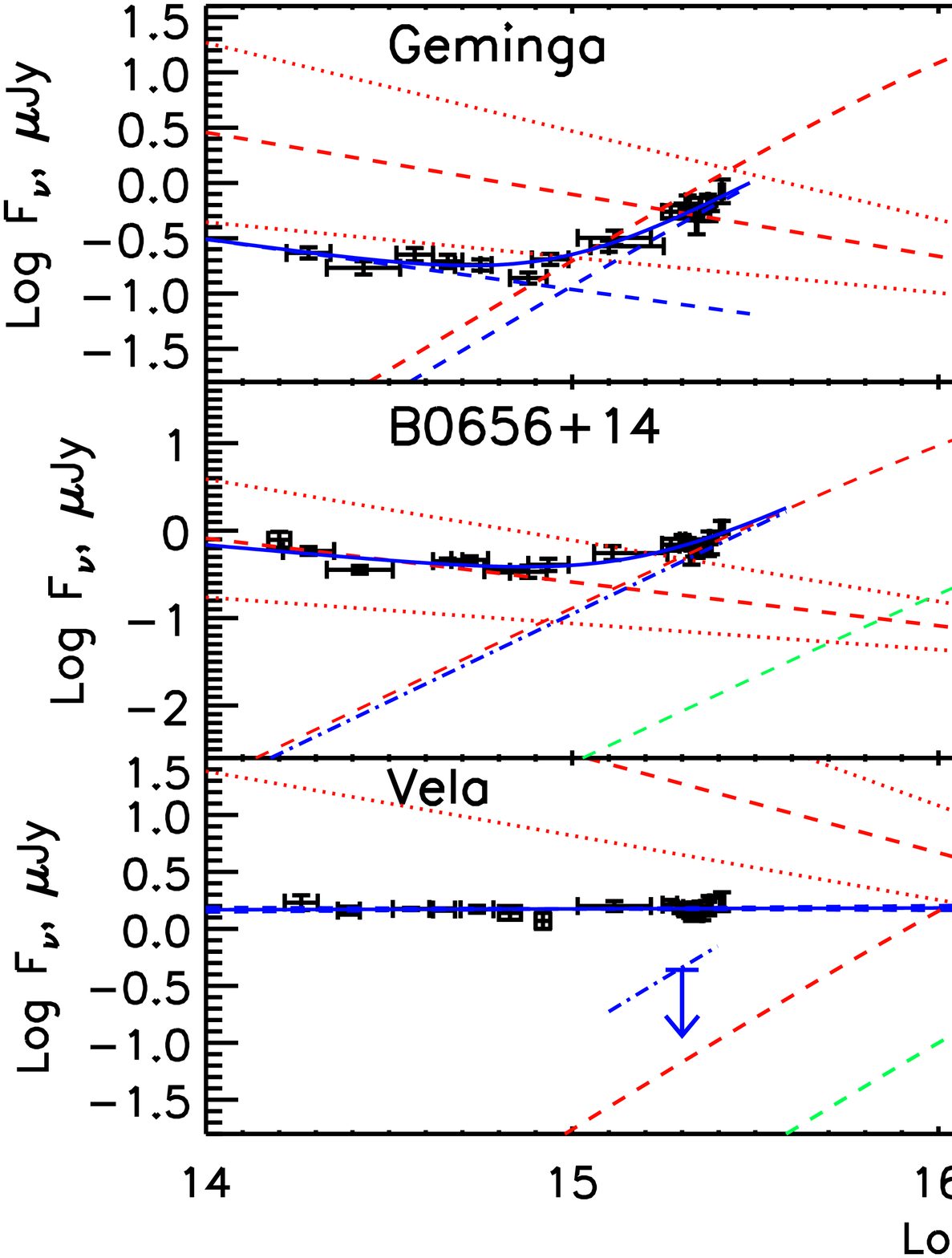}
\vspace{0.3cm}
 \caption{Unabsorbed multiwavelength spectra of the
Geminga, B0656+14 and Vela pulsars (top to bottom). In all cases the
X-ray spectra
(thick red lines) are fitted
with the TS+TH+PL model. The TS and PL X-ray components are shown
with dashed red lines
(the uncertainties of the PL fit are
dotted lines). The TH component is shown by the green dashed
line.
The solid blue lines show PL+BB fits to the optical
through FUV spectra; the blue dashed and dash-dotted lines
are the contributions of the PL and BB components, respectively.
}
\label{fig:4}       
\end{figure*}

\section{Multiwavelength spectra and light curves}

To compare the UV emission
 with the NIR, optical and X-ray emission,
we plotted multiwavelength spectra of the three pulsars
 in Figure 1 (NIR to FUV) and Figure 3 (NIR to X-rays). The FUV, NUV and X-ray light curves are
shown in Figure 2. Below we briefly discuss the multiwavelength
properties for each of the pulsars.

\subsection{Geminga}

 The top panel of Figure 1 shows the UV fluxes together with
 the multi-band NIR-optical photometry adopted from
Kargaltsev et al.\ (2005) and Shibanov et al.\ (2006).

As the NIR through FUV spectrum
cannot be described by a simple PL
model,
we
fit this spectrum with a
BB+PL model.
 For the fixed $R/d = 13\,{\rm km}/200\,{\rm pc}$, we obtain
$T=0.30 \pm 0.02$ MK, $\alpha_{\rm O} = - 0.46\pm 0.12$, $F_0 = 0.11\pm
0.02\,\mu{\rm Jy}$ for $E(B-V)=0.03$,
 where $\alpha_{\rm O}$
and $F_0$ are the parameters of the PL component: $F_\nu = F_0\,
(\nu/1\times 10^{15}\,{\rm Hz})^{\alpha_{\rm O}}$.
Notice that the parameters of
the BB component are virtually the same as obtained from the
FUV-MAMA spectrum alone.
 We see that the BB
emission dominates at $\lambda \lesssim 3000$ \AA, while the
 magnetospheric PL emission dominates at longer wavelengths.

To investigate the connection between the optical/UV and X-ray
properties, we use the data obtained with XMM-Newton. The X-ray
spectrum can be reasonably well described by a three-component
TS+TH+PL model (see \S1 and Kargaltsev et al.\ 2005 for details),
shown in the top panel of Figure 3.  One can see that the
extrapolation of the TS component {\em overpredicts} the FUV fluxes
by a factor of 1.6 [for $E(B-V)=0.03$; see also Fig.\ 1]. This FUV
{\em deficit} can also be demonstrated by plotting together the
 temperature-radius confidence contours for the TS component and for
   the BB fit to the FUV-MAMA spectrum (Fig.\ 4). We see
that at plausible values of interstellar extinction,
$E(B-V)\lesssim 0.07$,
 the FUV contours lie at smaller radii
(or much lower temperatures) than the X-ray contours,
in contrast to
 some other neutron stars (see \S4.2).
The extrapolation of the X-ray PL component
suffers from large uncertainties and
is marginally consistent with the
optical fluxes.

The background-subtracted X-ray light curves are compared with the
UV and optical light curves in Figure 2. In the 0.2--0.5 and
2--8 kev bands, the radiation is dominated by the TS and PL
components, respectively
  The
PL light
curve (pulsed fraction $f_{\rm p}=34\%\pm 8\%$) shows two pronounced
peaks per period, resembling the $\gamma$-ray light curve
(Jackson et al.\ 2002), albeit
with a smaller distance between the peaks, and a hint of a third
peak, at $\phi \approx 0.2$. On the contrary, the 0.2--0.5 keV light curve
($f_{\rm p} = 30\% \pm 2\%$) is characterized by one broad peak per
period (with small ``ripples'', perhaps due to contribution from the
PL and TH components).
 The
minimum of the TS light curve is approximately aligned in phase with
one of the minima of the PL light curve, being shifted by
$\Delta\phi \approx 0.1$ from the sharp dips of the NUV and FUV
light curves.

As the thermal soft emission dominates the spectrum from 4 eV
to 1 keV,
the non-thermal pulsations can only be observed in
the optical or hard X-rays. The double-peaked shape of the
hard X-ray (2--8 keV) pulsations
does not correlate with the TS
pulsations.
 The optical pulsations, marginally detected in
the B band (Shearer et al.\ 1998), also show a double-peaked structure,
with a larger separation between the peaks ($\Delta\phi\approx 0.6$
in the optical vs.\ $\Delta\phi\approx 0.4$ in the hard X-rays).
 The phases of the optical peaks differ from those
of the hard X-ray peaks. To understand and resolve these apparent
inconsistencies, a higher quality optical light curve should be
obtained.

\subsection{B0656+14}

Similar to Geminga,
we fit the NIR through FUV spectrum of B0656 (middle
panel in Fig.\ 1; NIR and optical data are from
Koptsevich et al.\ 2001 and Shibanov et al.\
2006)
with a two-component, BB+PL,
model.
 For the fixed $R/d = 13\,{\rm km}/288\,{\rm pc}$, we found the
 temperature
$T=0.50\pm0.05$ MK, spectral index  $\alpha_{\rm O} = -0.41\pm0.08$,
 and PL normalization $F_0 = 0.27\pm0.02\,\mu{\rm Jy}$ for $E(B-V)=0.02$.
One can see that the BB emission dominates at $\lambda \lesssim
2000$ \AA.
The fit also confirms that, in the NUV,
 the fraction of thermal photons is smaller than the fraction of
nonthermal photons by a factor of 1.5, which is compatible with the
deeper minima and higher pulsed fraction in the NUV light
curve.

We plot the NIR to X-rays spectra for B0656
in the middle panel of Figure 3. The  X-ray spectrum is again fitted
with the TS+TH+PL model.
We see that the extrapolation of the TS component from X-rays to
lower frequencies roughly coincides with the
FUV spectral flux
 and
goes slightly above (by a factor of 1.3) the UV thermal component.
This excess, however, is not statistically significant
(see
Fig.\ 4).

\begin{figure}
\centering
\vbox{
  \includegraphics[width=3.1in,angle=0]{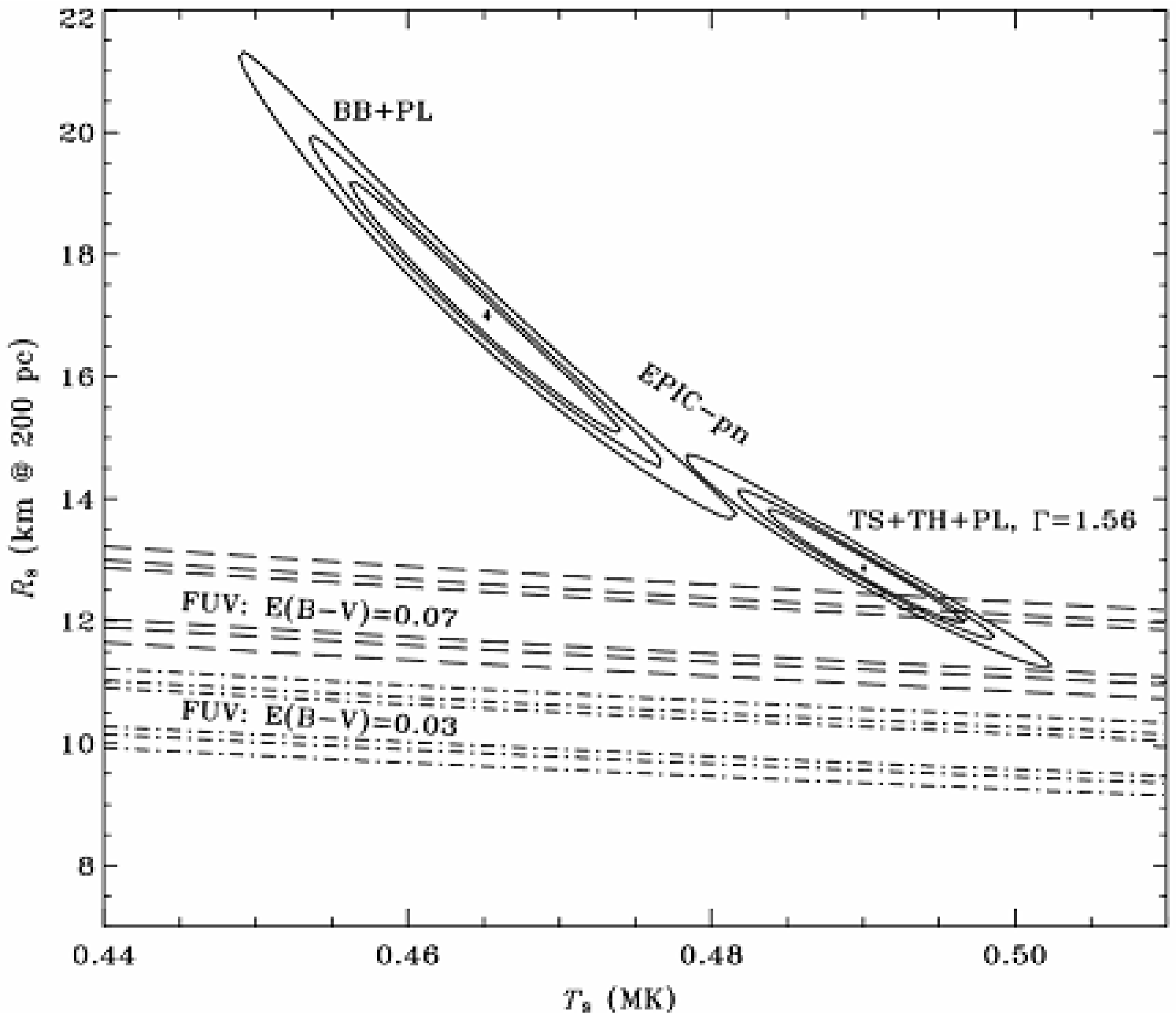}\hspace{-0.5cm}
\includegraphics[width=2.4in,angle=90]{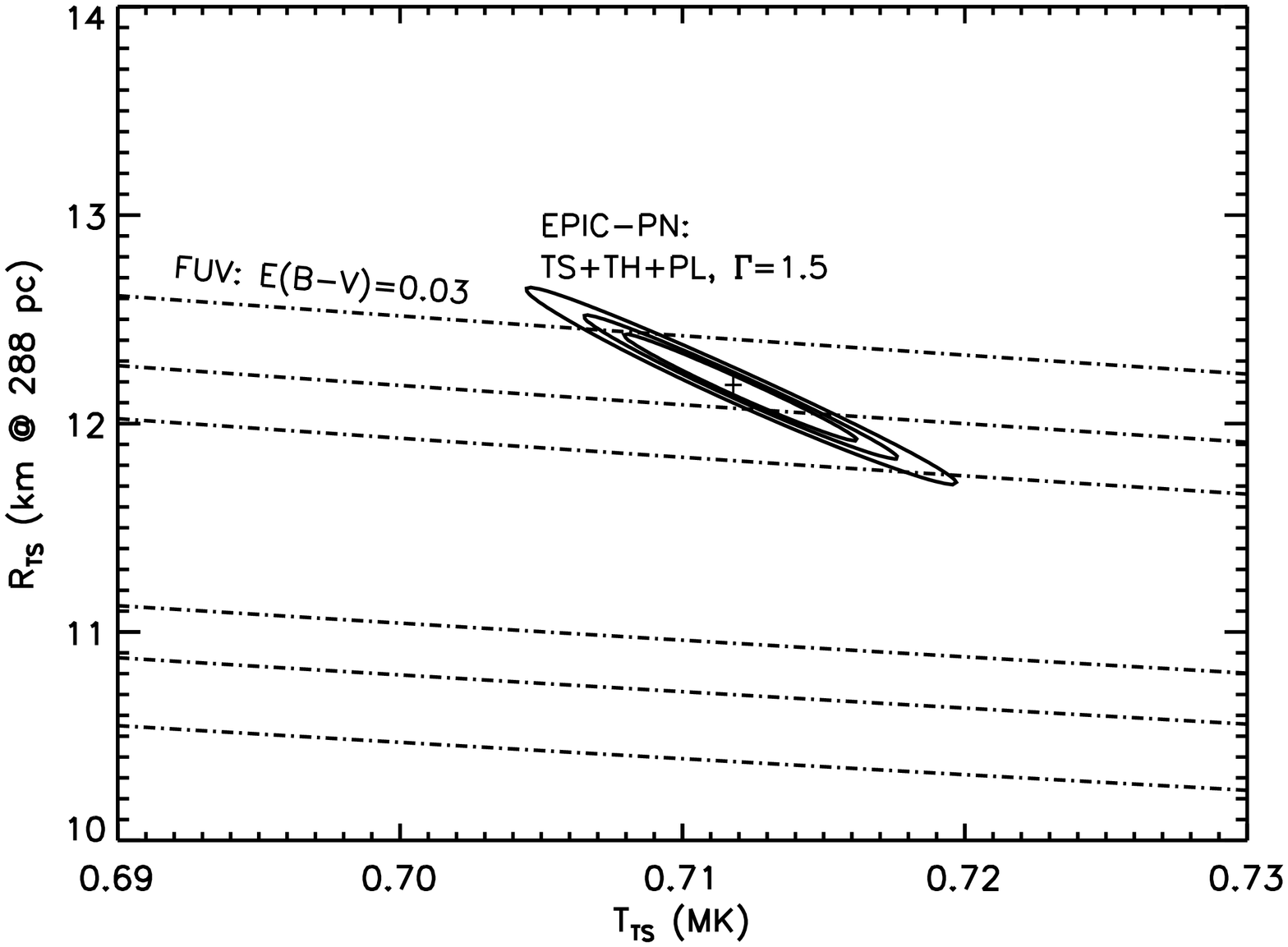}}
\caption{{\em Top:} Confidence contours (68\%, 90\%, and 99\%)
 in the (soft) temperature-radius plane obtained
from fitting the EPIC-pn spectra (solid lines) with the TS+TH+PL and
BB+PL models (labels near the contours).
The dashed and dash-dotted lines show the confidence contours
obtained from fitting the FUV spectrum with a BB model for two
values of $E(B-V)$. See Kargaltsev et al.\ (2005) for other details.
{\em Bottom:} Similar confidence contours but for B0656+14. }
\label{fig:3}       
\end{figure}

 The double-peaked structure seen in the optical (Kern et
 al.\ 2003) and NUV/FUV light curves
 changes over to a single peak in the X-ray light curves (Fig.\ 2).
It is possible that the asymmetric 0.25$-$0.7 keV pulse profile
consists of two components: a broad one, peaked at about
$\phi\approx0.60$--$0.65$, and a narrower component centered at
$\phi\approx0.90$--$0.95$, close to the radio pulse phase, $\phi=0$.
In this case the maximum of the broader component would
approximately coincide with the shallower minimum in the NUV and FUV
light curves, which is located at the phase of a presumable maximum
in the thermal UV component (see \S2.2). The peak of the narrow
component is approximately aligned with the single peak of the hard
(2$-$4 keV) X-ray light curve.

 Given the non-thermal interpretation
of the peaks in the optical-UV light curves and the close match
between the optical and X-ray PL components (Fig.\ 3), it is
surprising that the 2$-$4 keV pulse profile of B0656 shows only one
peak (at the phase of the second optical/UV peak;
$\phi\approx0.85$).
 Since the
phase-resolved spectroscopy shows similar PL slopes in
 both UV peaks,
 they are likely to
be of a similar
origin. The absence of the second peak at
the X-ray frequencies could be explained
assuming that the X-rays are
emitted closer to the NS in a more narrow cone while
the UV emission
is generated higher in the magnetosphere and spread over a
broader beam.
Alternatively,
it could be that the ``X-ray PL'' is not magnetospheric emission
(see \S4.3),
and the match between the X-ray and optical spectra
is just a
coincidence.

\subsection{Vela pulsar}

Figure 1 shows the spectrum of the Vela pulsar from NIR to UV, with
the optical/NIR data points adopted from Shibanov et al.\ (2003).
This very flat spectrum nicely fits a
PL model. The best-fit spectral index and normalization are
$\alpha_{\rm O}=0.01\pm0.02$ and $F_{\nu}=1.50\pm0.03$ $\mu$Jy (at
$\nu=10^{15}$ Hz), respectively. The spectral index is larger than
that obtained by Shibanov et al.\ (2003), $\alpha_{\rm O}=-0.12$,
 from fitting only IR and optical data.
On the other hand, the optical-UV spectrum is substantially flatter
than the non-thermal X-ray spectrum,
 $\alpha_X\approx-1$.

The UV spectrum of the Vela pulsar
allows one to obtain a restrictive upper limit on the NS
 surface temperature (see Romani et al.\ 2005 for details).
 The spectrum remains largely nonthermal even at the pulse minima,
 which constrains the NS temperature to $\lesssim0.4(d/300~{\rm pc})^{2}$
$(R/13~{\rm km})^{-2}$
 MK, assuming the BB model
  for the thermal emission
 and $E(B-V)=0.05$. The
 limit becomes less restrictive, $\lesssim0.6(d/300~{\rm pc})^{2}(R/13~{\rm km})^{-2}$ MK, if
 the hydrogen NS atmosphere
 model
is used to describe the thermal spectrum.
   The inferred upper limit
is surprisingly low compared to the temperature
of the older B0656 (Table 3).

In the UV, the Vela pulse profile is
dominated by multicomponent nonthermal emission.   In
 Figure 2 we show the NUV, FUV and X-ray light curves together. The X-ray light
curves obtained with  Chandra (Sanwal et al.\ 2002) and  RXTE
(Harding et al.\ 2002) are extremely complex, with multiple peaks
likely originating from different regions of magnetosphere or
corresponding to physically different spectral components.
 In agreement with
Harding et al.\ (2002), we identify at least five components in the
multiwavelength pulse profile. No peaks are seen in the UV at the
phases $\phi=0.12$ and 0.56 of the two $\gamma$-ray peaks (Kanbach
et al.\ 1994). Instead, in the FUV they are replaced by tighter
spaced peaks 1 and 2, which continue to the optical. We also find
that the UV peaks 2 and 4 have counterparts in the soft X-ray light
curve while peaks 1 and 3 appear to be shifted by
$\Delta\phi\approx0.1$ with respect to the other two soft X-ray
peaks. The UV light curves show strong correlation with the hard
X-ray light curves obtained with RXTE above 10 keV (Harding et al.\
2002), i.e.\ peaks 2, 3 and 4 are almost at the same phases in both
UV and  RXTE bands. At the same time, the correlation between the UV
and 2.5--8 keV light curves is less obvious. The spectral indices of
the individual UV components are rather uncertain and show no
correlation with the RXTE components (Romani et al.\ 2005). Should a
better NUV/FUV data be ever obtained, a joint phase-resolved
analysis of the UV and hard X-ray data may be more revealing.

\section{Discussion}

\subsection{Thermal emission from NS surface}

The spectra of all the three pulsars show thermal emission,
but the strength and position of the thermal hump vary. In the
Vela pulsar, thermal emission is seen only in soft (0.3--2 keV)
X-rays while in Geminga and B0656
 the FUV spectra are also predominantly thermal.
 The observed FUV and
soft X-ray spectra represent the
R-J and Wien tails
of the TS component. The TS BB temperatures measured in X-rays
decrease from $\approx 1.2$ MK in the Vela pulsar to $\approx0.7$ MK
in B0656 and $\approx0.5$ MK in the oldest Geminga. At the
same time, the bolometric TS luminosity of the Vela pulsar is
surprisingly low, which can be formally explained by a smaller
emitting area of the TS component. The actual reason for this
difference could be different properties (particulary, chemical
composition and degree of ionization) of the NS atmosphere. In both Geminga
and B0656 the brightness temperatures derived from the PL+BB fit to
the combined NIR through UV data are
lower than the
temperatures of the X-ray TS component
(see Table 3). The observed thermal UV spectrum of Geminga lies a factor
of 1.5$-$2 below the continuation of the X-ray TS component (for a
plausible extinction), whereas
 the UV deficit is
substantially smaller, if present at all, in B0656.
In the case of Vela pulsar, we could only estimate an upper limit on the
UV brightness temperature, which lies above the extrapolation of the X-ray
TS component.

Turning to the light curves (Fig.~2), one can see that in Geminga
and B0656 thermal soft X-ray light curves show broad
 maxima with some non-thermal ``bumps'' on top of them. The
 bump  (centered at $\phi\approx0.9$) is relatively large in B0656,
  making the soft X-ray light
curve
asymmetric. The soft X-ray pulse profile of Vela
is much more complex,
reflecting the larger
non-thermal contribution than in the other two pulsars. However, one
still can argue that there is a single broad thermal maximum in the
$\phi=0.7-1.2$ range.

The FUV light curves of B0656 and, especially, Gemin-ga
are also expected to be
largely thermal.
 Yet, Geminga shows strong double-peaked FUV pulsations,
different from both the NUV and the soft X-ray pulsations
(see Fig.~3). Neither FUV nor
soft X-ray pulsations of Gemin-ga can be produced by  locally
isotropic blackbody emission. Although the soft X-ray pulsations can
be
 explained by a
magnetized atmosphere model (e.g., Zavlin \& Pavlov 2002), the high
pulsed fraction and the peculiar shape of the FUV pulsations are not
predicted by this model (notice, however, that the current
atmosphere models are not applicable at the optical-UV frequencies).
A possible explanation of the strong
thermal pulsations invokes a
``screen'' of absorbing plasma suspended in the NS magnetosphere,
which may partially eclipse the surface emission at certain rotation
phases.

\subsection{UV emission of Geminga and B0656 versus RQINSs}

Over the last decade seven soft X-ray sources with thermal spectra
have been discovered by  ROSAT. Very large X-ray-to-optical flux
ratios/limits and a lack of radio emission allow one to conclude
 that these objects are radio-quiet isolated neutron
 stars (RQINSs). Extensive X-ray observations of RQINSs with  Chandra and
  XMM-Newton have provided high-quality spectra and period
 measurements for most of them (F.\ Haberl, these proceedings).
Recent period derivative measurements by Kaplan \& van Kerkwijk (2005ab)
 in RX~J0720.4-3125
and RX~J1308.6+2127
(J0720 and J1308 hereafter)
place these
RQINSs near the anomalous X-ray pulsars (AXPs) in the $P$--$\dot{P}$
diagram. However, the X-ray properties of RQINSs are quite different
from those of AXPs (the BB temperatures of RQINSs are much lower, and no PL
tails are seen). On the other hand, the X-ray spectra of
RQINSs are also very different from those of ordinary old radio
pulsars ($\tau\gtrsim 1$ Myr;  $B\sim 10^{12}$ G;
e.g., Zavlin \& Pavlov 2004b;
Kargaltsev et al.\ 2006).
In terms of
their thermal X-ray properties, RQINSs
resemble
middle-aged pulsars (see Table 3); however, the
optical-UV properties of RQINSs show important differences.

Out of the seven
RQINSs,
five have been detected
 in the optical, and UV spectra have been obtained for two of the five,
RX~J1856.5--3754 (J1856) and J0720 (van Kerkwijk \& Kulkarni
 2001; Kaplan et al.\ 2003). The UV
spectra of these RQINSs
look
thermal ($F_\nu \propto \nu^2$),
and even their optical fluxes seem to follow the same R-J curves,
although Kaplan et al.\ (2003) report
 a faint non-thermal PL compontent, with $\alpha_{\rm O}\approx 0.3$,
$L_{\rm PL,O}\sim 6\times 10^{27}$ ergs s$^{-1}$ for J0720
(for the other three RQINSs, the spectral slopes in the optical are
 uncertain).
We note that no detectable optical magnetospheric component is
expected at least in J0720 and J1308 (and likely in other RQINSs,
whose spin-down powers, $\dot{E}$, have not been measured yet)
 if the nonthermal optical efficiency of these objects is similar
to those of radio pulsars, $L_{\rm PL,O}/\dot{E}\sim
10^{-7}$--$10^{-6}$ (Zavlin \& Pavlov 2004b). For instance, one
would expect $L_{\rm PL,O}\sim 10^{24}$--$ 10^{25}$ ergs s$^{-1}$
for J0720, i.e. a factor of $>10^3$ higher efficiency is needed to
explain the result of Kaplan et al.\ (2003). More importantly,
RQINSs exhibit significant {\em excess in thermal UV emission}, with
the UV fluxes exceeding
 the continuations of the X-ray thermal spectra by a factor of 5--9.
 The nature of this UV
excess is not firmly established yet. It could occur if the UV and
soft X-ray thermal components are emitted from different regions
(with different areas) on a non-uniformly heated NS surface (Pavlov
et al.\ 2002).
Alternatively,
these NSs may have a condensed (solid) surface,
possibly covered by
a tenuous atmosphere that is optically thin in X-rays but optically
thick in the optical-UV
(e.g., Motch et al.\ 2003).
 Whatever is the nature of the large UV excess in RQINSs,
we do not see it in B0656 and see a {\em UV deficit} in Geminga.

The apparently smaller UV-emitting area of Geminga, as compared to
the X-ray-emitting area, cannot be explained by a nonuniform
temperature distribution. We might speculate that the temperature
distribution over the bulk of Geminga's surface is more uniform than
in J1856 and J0720,
e.g., because
of different geometry and strength of the magnetic field that
affects
the surface temperature
distribution. However, to explain why the more uniformly heated
Geminga exhibits quite substantial pulsations of its thermal
X-ray radiation
while no
pulsations have been detected from J1856,
one has to assume
very special orientations of the J1856's spin and magnetic axes.
The lack of UV excess in Geminga can also be attributed to different
chemical composition or lower temperature of the Geminga's surface
(Kargaltsev et al.\ 2005).
 However, the
latter argument does not apply to B0656 whose UV and soft X-ray
thermal components show higher temperatures, comparable to those of
J1856 and other RQINSs.
One could also speculate that the surface of RQINSs is
solid  while
in B0656 and Geminga it is in a gaseous state.
Indeed, the magnetic fields of Geminga and B0656 are 5--10 times
smaller than those of J0720 and J1308,
implying lower condensation
temperatures
(van Adelsberg et al.\ 2005).
In addition, irradiation of the NS surface by energetic particles and
photons generated in the magnetosphere may ablate the NS surface
and facilitate formation
of thick atmospheres in active pulsars while
we see no indications of magnetospheric activity in RQINSs.
The presence of thick atmospheres (lack of solid sufrace) in
pulsars is supported by the fact that
a hydrogen atmosphere model
provides a reasonably good description of the TS
spectrum in the Vela pulsar
while the BB model gives a very small NS radius
(Pavlov et al.\ 2001).
(Such fully ionized atmosphere models, however, are not directly applicable
to colder Geminga and B0656
because their atmospheres are not fully ionized.)
In RQINSs, the solid surface, directly seen in X-rays, would emit
the spectrum that resembles a BB but with the emissivity
 reduced by a factor of $\lesssim 2$ (van Adelsberg et al.\
2005), which would explain the apparently smaller X-ray radii of
RQINSs ($R_{\rm TS}$ in Table 3).
However, even a tenuous
atmosphere on top of the solid surface can be opaque in UV,
resulting in a
larger UV emitting area.
Within this interpretation, one still needs to explain the
absorbtion features observed in the spectra of several RQINSs.
Although it was suggested that these features can be formed in a
hydrogen or helium atmosphere (see van Kerkwijk \& Kaplan, these
proceedings), they could also be due to the resonances at
various hybrid frequencies
 in the dense magnetized condensate (van Adelsberg et al.\ 2005).

\subsection{Nonthermal emission}

In Geminga and B0656, the PL components fitted to the hard X-ray and
optical spectra show similar slopes
$\alpha_{\rm X}\approx\alpha_{\rm O}\approx -0.5$. However, in B0656
the extrapolation of the X-ray PL matches the optical points much
better that in Geminga (see Fig.\ 3).
For the Vela pulsar, the slope of the
PL component is much steeper in X-rays,
$\alpha_{\rm X}\approx -1$, than
in the UV-optical,
$\alpha_{\rm O}\approx 0.0$. However, one should keep in mind that
the slopes of the X-ray PL components are measured in the narrow
2--8 keV band (Fig.\ 3) and hence are very uncertain because of a
large background at these energies and, possibly, because the
spectrum deviates from a simple PL. For instance,
a comptonized tail of thermal radiation could
 mimic the nonthermal PL component in the narrow 2--8 keV band. Measuring hard X-ray
spectra above $\gtrsim 10$ keV will help to better understand the
nature of the nonthermal emission in Geminga and B0656.

The 2--8 keV light curves show rather sharp and strong peaks for
all the three pulsars. If the 2--8 keV  emission is indeed due to
the comptonization of thermal emission, one needs to explain its
high anisotropy.
The number of hard
X-ray peaks varies from a single peak in B0656's light curve to
a double-peaked
structure in Geminga, and even more
complex multi-peak light curve in the Vela pulsar.

In the UV-optical, the Vela pulsar shows the strongest
 pulsations, with at least four narrow peaks.
In contrast to the Vela pulsar, the
UV light curves of B0656 have a large thermal contribution
($\approx70$\% in FUV and $\approx50$\% in NUV).
Its NUV light curve shows two large peaks with a much wider
separation than the main peaks in the NUV pulse profile of the Vela
pulsar. Due to a much lower S/N in the B0656 NUV light curve,
smaller and narrower non-thermal peaks may remain undetected.
Similar to the Vela pulsar, the NUV light curve
 of B0656 is noticeably different from its 2--4 keV light curve.
 No obvious non-thermal contribution to the UV light curves is seen
 in Geminga, which is consistent with a large thermal fraction
 in the UV. However, non-thermal pulsations should be better seen
in the optical
($\lambda \lesssim 4300$ \AA).

\begin{table*}[t]
\caption{X-ray, UV and optical properties of middle-aged pulsars and
RQINSs.} \centering
\label{tab:3}       
\begin{tabular}{ccccccccccccc}
\hline\noalign{\smallskip}
 Name & $ T_{\rm UV}$ & $\log L_{\rm FUV}$ & $\log L_{\rm PL,O}$ & $\alpha_{\rm O}$ & $T_{ \rm TS}$ & $R_{\rm TS}$ & $\log L_{\rm TS}$  & $T_{\rm TH}$ & $R_{\rm TH}$ & $\log L_{\rm TH}$ & $\log L_{{\rm PL}, X}$ &  $\alpha_{\rm X}$  \\ [3pt]
           &    MK            & ergs/s   &    ergs/s   &            &   MK        &   km        &  ergs/s   & MK  & km   & ergs/s   &   erg/s    &              \\ [3pt] \tableheadseprule\noalign{\smallskip}
     Vela  & $\!\!\!\!\!\!\lesssim0.4$ & $29.15$  &   $28.57$   &  $\!\!-0.01$   &   $1.17$    &  $~3.7$      &  $32.26$  & 2.1 & 0.7  & 31.80    &   $31.49$  &  $\!\!-1.0$      \\
B0656+14   &    $0.53$        & $28.74$  &   $27.87$   &  $\!\!-0.41$   &   $0.71$    &  $12.2$     &  $32.43$  & 1.4 & 1.1  & 31.54    &   $31.15 $ &  $\!\!-0.5$      \\
Geminga    &    $0.30$        & $28.33$  &   $27.17$   &  $\!\!-0.46$   &   $0.49$    &  $12.9$     &  $31.83$  & 2.3 & ~~0.05 & 29.64    &   $30.34$  &  $\!\!-0.6$     \\
\hline\noalign{\smallskip}
RX~J0720  &    $0.37$        & $28.44$  &   $~~27.78?$   &  $~~0.3?$   &   $\!\!1.0$    &  $~5.1$     &  $32.58$  & ... & ... & ...    &   ...  &  ...     \\
RX~J1856  &    $0.80$        & $28.76$  &   ...   &  ...   &   $0.72$    &  $~8.7$     &  $32.65$        & ... & ... & ...  &   ...  &  ...     \\
 \noalign{\smallskip}\hline
\end{tabular}\\
\tablecomments{ First column is the brightness temperature measured
from the TS+PL fit to the NIR through FUV spectra (see Fig.\ 1) for
a 13 km NS radius. Second column gives the unabsorbed FUV luminosity
in the 1150--1700 \AA\ passband [$E(B-V)=0.05$, 0.03 and 0.03 for
Vela, B0656 and Geminga, respectively].  Third and  forth columns
are the PL component slope and luminosity (in 4000-9000 \AA),
respectively. Columns three through ten list the parameters of the
TS+TH+PL fits to the X-ray spectra. Thermal component luminosities
are the bolometric luminosities
 while the
unabsorbed PL
 luminosity, $L_{\rm PL,X}=4\pi
d^{2}\mathcal{F}_{\rm PL,X}$, is in the 0.2--10 keV band. The luminosities and
radii are calculated for the distances listed in Table 1.
 }
\end{table*}

\section{Conclusions}
\label{sec:concl}

We have observed
three pulsars (Geminga, B0656+14, and Vela)
at the far-UV and near-UV wavelengths with
the STIS MAMA detectors and measured the pulsar spectra and pulse
profiles. We have also analyzed the X-ray data and found that
optical through X-ray spectra consist of thermal and non-thermal
components in all the three pulsars. In particular, we found
the following.

 {\bf 1.}  Thermal contribution to the FUV passband increases
 with
pulsar age.
The thermal component is not seen in the
younger Vela pulsar ($\tau\approx11$ kyrs), but it dominates the FUV
spectrum in the Geminga pulsar ($\tau\approx340$ kyrs). As the FUV
spectrum grows more thermal, the ``thermal hump'', whose maximum is
located in soft X-rays/EUV, shifts toward lower frequencies.

 {\bf 2.} The X-ray spectra of the two older pulsars, Gemin-ga and B0656, resemble those of RQINSs.
 However, their UV properties are noticeably different. In sharp
contrast to RQINSs, whose spectra show large {\em UV/optical
excess}, Geminga shows {\em UV deficit} while the continuation of
the X-ray TS component approximately matches the R-J component in
B0656. Likely, this
 reflects the differences in the magnetic field and phase
 state of the NS surface layers.

 {\bf 3.} While no
UV pulsations have been detected in RQINSs,
Geminga shows strong,
non-sinusoidal pulsations in the FUV range, where the spectrum is
 dominated by the thermal component.
 To explain
  the
 strong FUV pulsations,
 one may need to invoke magnetospheric
 absorption
  at certain rotation phases.

 \begin{acknowledgements}
We thank Slava Zavlin for the help with the X-ray data analysis and
David Kaplan for providing the most recent parallax measurements for
RQINSs. This work was supported by STScI grants GO-9182 and GO-9797
and NASA grant NAG5-10865.
\end{acknowledgements}

\end{document}